\documentclass[12pt]{amsart}

\usepackage{array}
\usepackage{graphics}
\usepackage{epsfig}
\usepackage{lscape}
\usepackage{epsfig}
\usepackage{etex}
\usepackage{amscd}
\usepackage{amsmath}
\usepackage{amsfonts}
\usepackage{amssymb}

\newtheorem{theorem}{\bf Theorem}

\newtheorem{remark}[theorem]{\bf Remark}

\newtheorem{corollary}[theorem]{\bf Corollary}

\newtheorem{example}{\bf Example}

\font\Bbb=msbm10

\def\P{\hbox{\Bbb P}}
\def\C{\hbox{\Bbb C}}

\title{\large Human blood genotypes dynamics}

\thanks{Funding for this research was provided by the grant of the Russian
Federation Government to support scientific research under the
supervision of leading scientist at Siberian Federal University,
no.~14.Y26.31.0006. The author was also supported by the grants of
the Russian Foundation for Basic Research no.~13-01-12417-ofi-m2
and 14-01-00544-a.}

\author{Timur Sadykov}
\address{Dept. of Computer Science and Mathematics,
    Russian Plekhanov University, Smolnaya~36, Moscow
    125993, Russia}

\begin{document}

\large

\maketitle

\begin{abstract}
The frequencies of human blood genotypes in the ABO and Rh systems
differ between populations. Moreover, in a given population, these
frequencies typically evolve over time. The possible reasons for
the existing and expected differences in these frequencies (such
as disease, random genetic drift, founder effects, differences in
fitness between the various blood groups etc.) are in the focus of
intensive research. To understand the effects of historical and
evolutionary influences on the blood genotypes frequencies, it is
important to know how these frequencies behave if no influences at
all are present. Under this assumption the dynamics of the blood
genotypes frequencies is described by a polynomial dynamical
system defined by a family of quadratic forms on the
17-dimensional projective space. To describe the dynamics of such
a polynomial map is a task of substantial computational
complexity.

We give a complete analytic description of the evolutionary
trajectory of an arbitrary distribution of human blood variations
frequencies with respect to the clinically most important ABO and
RhD antigens. We also show that the attracting algebraic manifold
of the polynomial dynamical system in question is defined by a
binomial ideal.

\end{abstract}

\section{Introduction}

Since the discovery of human blood groups in~1900, their
distributions in various countries and ethnicities have been
attracting attention of researchers \cite{Anstee2010}. Such
distributions vary a lot across the world
\cite{Kang1997,Ohashi2006,Sato2010} and, in general, evolve over
time~\cite{Novitski1976}. It is classically known that in the
absence of evolutionary influences the allele frequency of a
single trait achieves the Hardy-Weinberg equilibrium
\cite{Novitski1976} already in the second generation and then
remains constant~\cite{Bernstein1923}. Besides, blood groups
frequencies satisfy an algebraic
relation~\cite{BernsteinFelix1929}. The interplay between the
frequencies of all possible genotypes or phenotypes combinations
of a pair of genes (even though their frequencies are
uncorrelated) is however much more complex~\cite{Lyubich1971}. For
a random initial population, the frequencies of all possible
combinations of blood group and Rh factor phenotypes will only
stabilize after an infinitely long evolution.

By computing the linkage disequilibria between alleles one can, in
principle, find the frequency of any genotype in a given
generation~\cite{Lyubich1971}. However, finding explicit analytic
formulas for the evolution of genotypes frequencies and invariant
varieties of the polynomial dynamical system describing this
evolution is a problem of great computational complexity. In the
present paper we develop a symbolic solution technique which
allows us to give a closed form description of the evolutionary
trajectory. We show how the frequencies of human blood genotypes
(distinguished by both blood group and Rh factor variations) with
arbitrary initial distribution will evolve after any given number
of generations in a population where no blood genotype is favored
over another with respect to the ability to pass its genes to the
next generation.

Throughout the paper, we will be denoting the blood group traits
by A,B,O, and the Rh factor traits by H (positive) and h
(negative). The 18 human blood genotypes in the ABO Rh system will
be denoted by OOhh (Rh negative 1st blood group), AOhh, AAhh,
BOhh, BBhh, ABhh, OOHh, AOHh, AAHh, BOHh, BBHh, ABHh, OOHH, AOHH,
AAHH, BOHH, BBHH, and ABHH (homozygously Rh positive 4th blood
group). In the sequel, we will always be using this particular
ordering of the blood genotypes. Since A and B traits are
codominant over O while the H trait is dominant over h, the above
genotypes comprise the 8 blood phenotypes: Oh (Rh negative 1st
blood group, same as OOhh), Ah (Rh negative 2nd blood group
comprising genotypes AOhh and AAhh), Bh, ABh, OH, AH (Rh positive
2nd blood group comprising genotypes AOHh, AOHH, AAHh, AAHH), BH,
and ABH.

In demography and transfusiology, it is often important to know
and predict the frequencies of blood genotypes or phenotypes with
respect to {\it both} blood group and Rh factor
\cite{Anstee2010,Ohashi2006,Okada,Reilly2007}. For instance, one
would like to know the expected frequency of the Rh negative 4th
blood group after a given number of years in a certain population.
As we will see later, the convergence of the blood genotypes
frequencies towards the limit distribution is rather slow (in the
real time scale) for generic choice of their initial distribution.
For instance, Table~1 shows that for $p=\frac{1}{2}$ the frequency
of the OH genotype in Example~\ref{ex:exampleWithTable} below
after two generations is~0.109 while its equilibrium value
is~0.187. Moreover, the limit distribution might not be ever
reached for a particular real population because of migration and
evolutionary influences that affect the blood genotypes
frequencies \cite{Anstee2010,Lyubich1971}\,. The Hardy-Weinberg
result gives the equilibrium genotypes frequencies after an
infinitely long evolution and the Bernstein equation
\cite{BernsteinFelix1929,Novitski1976} relates these frequencies
in a population that is already at the equilibrium.

The purpose of the present paper is to fill the gap between an
initial distribution and the equilibrium state (that is in general
only achieved after an infinitely long evolution) by giving an
explicit closed form analytic formula for the frequencies
distribution. We describe the evolution of the frequencies of all
possible genotypes of human blood in the clinically most important
ABO and Rh blood group systems for an arbitrary initial
distribution of these frequencies and after any number of
generations.


\section{Polynomial Dynamical System Describing the Evolution of a Distribution of Blood
Genotypes}

We will be assuming that blood group and Rh factor are
statistically correlated neither with gender nor with fertility or
any aspect of sexual behavior of a human. That is, we will
consider a population where an individual's chances to pass
her/his genes to the next generation do not depend on her/his
blood genotype.

Since we assume that blood genotypes frequencies are uncorrelated
with gender, their distribution is the same for males and females.
The blood genotype distribution in such a population is therefore
completely determined by a vector with 18 real nonnegative
components $x=(x_1,x_2,\ldots,x_{18}),$ where $x_1$ denotes the
frequency of OOhh, $x_2$ is the frequency of AOhh, etc. (see the
ordering of the blood genotypes introduced above). We will only
consider vectors not all of whose components are zero since zero
population has trivial dynamics. Moreover, since we are only
interested in the proportions of the population having prescribed
blood genotypes, we will identify proportional vectors. Thus for
the purpose of studying blood genotypes dynamics a population is
identified with a point in the 17-dimensional projective
space~$\P^{17}.$

Let $x=(x_1,x_2,\ldots,x_{18})\in\P^{17}$ be the vector encoding
the blood genotypes distribution in a population. Using the
well-known blood inheritance rules \cite{Hoffman2012} together
with the above statistical assumptions on the population under
study, we conclude that the distribution of blood genotypes in the
next generation is described by the vector whose components are
the following 18 quadratic forms:

{\tiny
\begin{equation} \hskip-0.9cm
\begin{array}{lcc}
f_1(x) &=& (4 x_1+2 x_2+2 x_4+2x_7+x_8+x_{10})^2,\\
f_2(x) &=& 2(4 x_1+2 x_2+2 x_4+2 x_7+x_8+x_{10})
(2 x_2+4 x_3+2 x_6+x_8+2 x_9+x_{12})\\
f_3(x) &=&  (2 x_2+4 x_3+2 x_6+x_8+2
x_9+x_{12}){}^2,\\
f_4(x) &=& 2 (4 x_1+2 x_2+2 x_4+2 x_7+x_8+x_{10}) (2 x_4+4 x_5+2
x_6+x_{10}+2
x_{11}+x_{12}),\\
f_5(x) &=&  (2 x_4+4 x_5+2x_6+x_{10}+2
x_{11}+x_{12}){}^2,\\
f_6(x) &=& 2 (2 x_2+4 x_3+2 x_6+x_8+2 x_9+x_{12})
(2 x_4+4 x_5+2 x_6+x_{10}+2 x_{11}+x_{12}),\\
f_7(x) &=& 2 (4 x_1+2 x_2+2 x_4+2 x_7+x_8+x_{10})
( 2 x_7+x_8+x_{10}+4 x_{13}+2 x_{14}+2 x_{16}),\\
f_8(x) &=& 4 (2 x_2 x_7+4 x_3 x_7+2 x_6 x_7+2 x_1
x_8+2 x_2x_8+2 x_3 x_8+x_4 x_8+x_6 x_8+2 x_7 x_8+x_8^2+4 x_1 x_9+ \\
&& 2x_2x_9+2 x_4 x_9+4 x_7 x_9+2 x_8 x_9+x_2 x_{10}+2 x_3 x_{10}+
x_6
x_{10}+x_8 x_{10}+2 x_9 x_{10}+2 x_1 x_{12}+x_2 x_{12}+\\
&& x_4x_{12}+2 x_7 x_{12}+x_8 x_{12}+x_{10} x_{12}+4 x_2 x_{13}+8
x_3
x_{13}+4 x_6 x_{13}+2 x_8 x_{13}+4 x_9 x_{13}+2 x_{12} x_{13}+\\
&& 4x_1 x_{14}+4 x_2 x_{14}+4 x_3 x_{14}+2 x_4 x_{14}+2 x_6
x_{14}+2
x_7 x_{14}+2 x_8 x_{14}+2 x_9 x_{14}+x_{10} x_{14}+x_{12}x_{14}+\\
&& 8 x_1 x_{15}+4 x_2 x_{15}+4 x_4 x_{15}+4 x_7 x_{15}+2 x_8
x_{15}+2
x_{10} x_{15}+2 x_2 x_{16}+4 x_3 x_{16}+2 x_6 x_{16}+\\
&& x_8x_{16}+2 x_9 x_{16}+x_{12} x_{16}+4 x_1 x_{18}+2 x_2
x_{18}+2
x_4 x_{18}+2x_7 x_{18}+x_8 x_{18}+x_{10} x_{18}),\\
f_9(x) &=& 2 (2 x_2+4 x_3+2 x_6+x_8+2 x_9+x_{12})
(x_8+2 x_9+x_{12}+2 x_{14}+4 x_{15}+2 x_{18}),\\
f_{10}(x) &=& 4 (2 x_4 x_7+4 x_5 x_7+2 x_6 x_7+x_4
x_8+2 x_5 x_8+x_6 x_8+2 x_1 x_{10}+x_2 x_{10}+2 x_4 x_{10}+2 x_5x_{10}+x_6 x_{10}+       \\
&& 2 x_7 x_{10}+x_8 x_{10}+x_{10}^2+4 x_1
x_{11}+2 x_2 x_{11}+2 x_4 x_{11}+4 x_7 x_{11}+2 x_8 x_{11}+2x_{10} x_{11}+2 x_1 x_{12}+         \\
&& x_2 x_{12}+x_4 x_{12}+2 x_7 x_{12}+x_8
x_{12}+x_{10} x_{12}+4 x_4 x_{13}+8 x_5 x_{13}+4 x_6 x_{13}+ 2x_{10} x_{13}+4 x_{11} x_{13}+     \\
&& 2 x_{12} x_{13}+2 x_4 x_{14}+4 x_5
x_{14}+2 x_6 x_{14}+x_{10} x_{14}+2 x_{11} x_{14}+x_{12}x_{14}+4 x_1 x_{16}+2 x_2 x_{16}+4 x_4 x_{16}+     \\
&& 4 x_5 x_{16}+2 x_6 x_{16}+2
x_7 x_{16}+x_8 x_{16}+2 x_{10} x_{16}+2 x_{11} x_{16}+x_{12}x_{16}+8 x_1 x_{17}+4 x_2 x_{17}+    \\
&& 4 x_4 x_{17}+4 x_7 x_{17}+2 x_8 x_{17}+2 x_{10} x_{17}+4 x_1
x_{18}+2 x_2 x_{18}+2 x_4x_{18}+2 x_7x_{18}+x_8 x_{18}+x_{10} x_{18}),\\
f_{11}(x) &=& 2 (2 x_4+4 x_5+2 x_6+x_{10}+2 x_{11}+x_{12})
(x_{10}+2 x_{11}+x_{12}+2 x_{16}+4 x_{17}+2 x_{18}),\\
f_{12}(x) &=& 4 (x_4 x_8+2 x_5 x_8+x_6 x_8+2 x_4 x_9+4 x_5 x_9+2
x_6 x_9+x_2 x_{10}+2 x_3 x_{10}+x_6
x_{10}+x_8x_{10}+2 x_9 x_{10}+   \\
&&2 x_2 x_{11}+4 x_3 x_{11}+2 x_6 x_{11}+2 x_8
x_{11}+4 x_9 x_{11}+x_2 x_{12}+2 x_3 x_{12}+x_4 x_{12}+2 x_5x_{12}+2 x_6 x_{12}+  \\
&& x_8 x_{12}+ 2 x_9 x_{12}+x_{10} x_{12}+2
x_{11} x_{12}+x_{12}^2+2 x_4 x_{14}+4 x_5 x_{14}+2 x_6 x_{14}+x_{10} x_{14}+2 x_{11} x_{14}+       \\
&& x_{12} x_{14}+4 x_4 x_{15}+8 x_5 x_{15}+4 x_6 x_{15}+2 x_{10} x_{15}+4 x_{11} x_{15}+2 x_{12}x_{15}+2 x_2 x_{16}+4 x_3 x_{16}+2 x_6 x_{16}+    \\
&& x_8 x_{16}+2 x_9x_{16}+x_{12} x_{16}+4 x_2 x_{17}+8 x_3 x_{17}+4 x_6 x_{17}+2 x_8x_{17}+4 x_9 x_{17}+2 x_{12} x_{17}+2 x_2 x_{18}+      \\
&&4 x_3 x_{18}+2 x_4x_{18}+4 x_5 x_{18}+4 x_6 x_{18}+x_8
x_{18}+2x_9 x_{18}+x_{10}x_{18}+2x_{11} x_{18}+2 x_{12} x_{18}),\\
f_{13}(x) &=&  (2 x_7+x_8+x_{10}+4 x_{13}+2 x_{14}+2
x_{16}){}^2,\\
f_{14}(x) &=& 2 (2 x_7+x_8+x_{10}+4 x_{13}+2 x_{14}+2 x_{16})
(x_8+2 x_9+x_{12}+2 x_{14}+4 x_{15}+2
x_{18}),\\
f_{15}(x) &=&  (x_8+2 x_9+x_{12}+2 x_{14}+4 x_{15}+2
x_{18}){}^2,\\
f_{16}(x) &=& 2 (2 x_7+x_8+x_{10}+4 x_{13}+2 x_{14}+2 x_{16})
(x_{10}+2 x_{11}+x_{12}+2 x_{16}+4
x_{17}+2 x_{18}),\\
f_{17}(x) &=& (x_{10}+2
x_{11}+x_{12}+2 x_{16}+4 x_{17}+2 x_{18}){}^2,\\
f_{18}(x) &=& 2 (x_8+2 x_9+x_{12}+2 x_{14}+4 x_{15}+2 x_{18})
(x_{10}+2 x_{11}+x_{12}+2 x_{16}+4 x_{17}+2 x_{18}).
\end{array}
\label{eq:explicitPolynomialDynamicalSystem}
\end{equation}
}
For instance, the polynomial $f_1(x)$ can be obtained by observing
that the only blood genotypes that contribute to the frequency of
the genotype OOhh in the next generation are OOhh, AOhh, BOhh,
OOHh, AOHh, and BOHh. Recall that their frequencies in the initial
generation are denoted by $x_1, x_2, x_4, x_7,$ $x_8,$ and
$x_{10},$ respectively. In a family where both parents' blood
belongs to the OOhh genotype, 100\% of the children will have the
same blood. An offspring of the parents with the blood genotype
AOhh will have blood of the type OOhh with the probability 1/4.
Computing the probabilities for an offspring to have blood of the
type OOhh for all possible combinations of the parents' blood
genotypes and clearing common denominators (this makes use of our
projective model and must be done for all components of the
polynomial map (\ref{eq:explicitPolynomialDynamicalSystem})
simultaneously), we arrive at $f_1(x).$ The other components
of~(\ref{eq:explicitPolynomialDynamicalSystem}) are obtained by
means of similar arguments.

The polynomials (\ref{eq:explicitPolynomialDynamicalSystem}) vary
greatly in their complexity and three patterns are easily
distinguishable. The polynomials that are squares of linear forms,
that is, $f_1, f_3, f_5,$ $f_{13}, f_{15}, f_{17},$ correspond to
blood genotypes that are homozygous for both blood group and Rh
factor. The polynomials that are products of two different linear
forms, that is, $f_2, f_4, f_6, f_7, f_9, f_{11},$ $f_{14},
f_{16}, f_{18},$ correspond to the blood genotypes that are
homozygous for either blood group or Rh factor but not both.
Finally, the three complicated polynomials $f_8, f_{10}, f_{12}$
are the counterparts of the fully heterozygous genotypes AOHh,
BOHh, and ABHh.

Observe that no particular population growth model has been used
for computing the polynomials $f_1,\ldots,f_{18}$ since our goal
is to compute the frequencies of the blood genotypes in the next
generation no matter how numerous it is. (It only has to be
numerous enough for the law of large numbers to hold.) Choosing a
particular growth model would result in multiplying the
polynomials $f_1,\ldots,f_{18}$ with a common normalizing
function.

Thus the distribution of the blood genotypes in the next
generation is completely described by the polynomial vector-valued
function $f(x) = f_1(x),\ldots,$ $f_{18}(x)$ from the projective
space $\P^{17}$ into itself:
$$
\begin{array}{rcl}
f : \P^{17} & \rightarrow & \P^{17}, \\
f : x  =  (x_1,\ldots,x_{18}) & \mapsto &
(f_1(x),\ldots,f_{18}(x)).
\end{array}
$$
Such a map defines a polynomial dynamical system. Finding blood
genotypes distributions in subsequent generations means computing
the sequence of iterates
$$
f(x), \, f(f(x)), \ldots, f^n(x) =f(f(\ldots f(x) \ldots )),\ldots
$$
of the polynomial vector-valued function~$f.$ Here $f^n(x)$ is
what the initial distribution $x$ evolves into after $n$
generations.

Typically a polynomial dynamical system on a complex manifold does
not admit an explicit analytic description of the trajectory of a
generic point. The vast majority of the results in complex
dynamics are ergodic-theoretic in nature
\cite{BedfordJonsson2000,Cantat2010}. However, the biological
origin of the dynamical
system~(\ref{eq:explicitPolynomialDynamicalSystem}) suggests that
it should not exhibit any chaotic behavior.

The polynomial dynamical
system~(\ref{eq:explicitPolynomialDynamicalSystem}) is the main
object of study in the paper. We aim to find an explicit symbolic
description of the orbit of any initial distribution of human
blood genotypes frequencies under the action
of~(\ref{eq:explicitPolynomialDynamicalSystem}) and to describe
its rate of convergence towards the equilibrium.

It will often be convenient to identify a distribution of blood
genotypes in a population with a linear form whose coefficients
are proportions of people with given blood genotypes and whose
formal variables are the 18 blood genotypes names OOhh, AOhh,
\ldots, ABHH. For instance, the linear form $p\cdot$OOhh +
$(1-p)\cdot$ABHH denotes the population where $p$ people have
blood genotype OOhh and $1-p$ people have genotype ABHH.


\section{Explicit symbolic description of evolutionary trajectories}

The main result of the paper is the following statement.
\begin{theorem}
\label{thm:mainTheorem} The initial distribution of blood
genotypes frequencies $x=(x_1,x_2,$ $\ldots,x_{18}) \in\P^{17}$
will after~$n\geq 1$ generations evolve into the distribution
{\small
\begin{equation}
\begin{array}{l}
           f^n(x)= Q\left(\right.  M_1 (M_5 + M_6) -  M_4 (M_2 + M_3) + 2^{n-1} (M_1 + M_4) (M_1 + M_2 + M_3), \\
\quad \phantom{f^n(x)=\left(\right. }  M_2 (M_4 + M_6) -  M_5 (M_1 + M_3) + 2^{n-1} (M_2 + M_5) (M_1 + M_2 + M_3), \\
\quad \phantom{f^n(x)=\left(\right. }  M_3 (M_4 + M_5) -  M_6 (M_1 + M_2) + 2^{n-1} (M_3 + M_6) (M_1 + M_2 + M_3), \\
\quad \phantom{f^n(x)=\left(\right. }  M_4 (M_2 + M_3) -  M_1 (M_5 + M_6) + 2^{n-1} (M_1 + M_4) (M_4 + M_5 + M_6), \\
\quad \phantom{f^n(x)=\left(\right. }  M_5 (M_1 + M_3) -  M_2 (M_4 + M_6) + 2^{n-1} (M_2 + M_5) (M_4 + M_5 + M_6), \\
\quad \phantom{f^n(x)=\left(\right. }  M_6 (M_1 + M_2) -  M_3 (M_4
+ M_5)+  2^{n-1} (M_3 + M_6) (M_4 + M_5 + M_6) \left. \right).
\end{array}
\label{eq:bloodDistrAfterNGenerations}
\end{equation}
}
Here $M_i = M_i(x)$ is the $i$-th component of the image of the
initial distribution $x\in\P^{17}$ under the linear map $M:\P^{17}
\rightarrow \P^{5}$ defined by the matrix
\begin{equation}
\phantom{-------}
M = \left(
\begin{array}{cccccccccccccccccc}
 4 & 2 & 0 & 2 & 0 & 0 & 2 & 1 & 0 & 1 & 0 & 0 & 0 & 0 & 0 & 0 & 0 & 0 \\
 0 & 2 & 4 & 0 & 0 & 2 & 0 & 1 & 2 & 0 & 0 & 1 & 0 & 0 & 0 & 0 & 0 & 0 \\
 0 & 0 & 0 & 2 & 4 & 2 & 0 & 0 & 0 & 1 & 2 & 1 & 0 & 0 & 0 & 0 & 0 & 0 \\
 0 & 0 & 0 & 0 & 0 & 0 & 2 & 1 & 0 & 1 & 0 & 0 & 4 & 2 & 0 & 2 & 0 & 0 \\
 0 & 0 & 0 & 0 & 0 & 0 & 0 & 1 & 2 & 0 & 0 & 1 & 0 & 2 & 4 & 0 & 0 & 2 \\
 0 & 0 & 0 & 0 & 0 & 0 & 0 & 0 & 0 & 1 & 2 & 1 & 0 & 0 & 0 & 2 & 4 & 2
\end{array}
\right) \label{matrixM}
\end{equation}
while $Q$ is the quadratic map $Q:\P^5 \rightarrow \P^{17}$
defined by
\begin{equation}
\begin{array}{l}
Q(y_1, y_2, y_3, y_4, y_5, y_6) = \\
\phantom{--------}\phantom{-} \left(y_{1}^2, \, 2 y_{1} y_{2}, \, y_{2}^2, \, 2y_{1} y_{3}, \, y_{3}^2, \, 2 y_{2} y_{3}, 2 y_{1} y_{4}, \right. \\
\phantom{--------}\phantom{--} 2 y_{1} y_{5} + 2 y_{2} y_{4}, \, 2 y_{2} y_{5}, \, 2 y_{1} y_{6} + 2 y_{3} y_{4}, \, 2 y_{3} y_{6}, \\
\phantom{--------}\phantom{--} 2 y_{2} y_{6} + 2 y_{3} y_{5}, \,
y_{4}^2, \, 2y_{4} y_{5}, \, y_{5}^2, \, 2y_{4} y_{6}, \, y_{6}^2,
\, 2 y_{5} y_{6}\left.\right).
\end{array}
\label{mapQ}
\end{equation}
\end{theorem}

\begin{remark}\rm
Although $n$ stands for the integer number of iterations of the
polynomial map $f,$ the formula
(\ref{eq:bloodDistrAfterNGenerations}) makes perfect sense for any
real $n\geq 1.$ As we will see later in
Section~\ref{sec:discussion}, it provides a smooth interpolation
of the blood genotypes frequencies in subsequent generations
described by our discrete model.
\end{remark}

\begin{remark}\rm
Having the explicit expression
(\ref{eq:bloodDistrAfterNGenerations}) for the $n$th iterate of a
polynomial map~(\ref{eq:explicitPolynomialDynamicalSystem}), it is
tempting to try to prove it by induction. Let $R(x,n)$ denote the
right-hand side of~(\ref{eq:bloodDistrAfterNGenerations}). It is
easy to check that $R(x,1)\equiv f(x)$ in $\P^{17}$ (that is,
these two polynomial vectors are proportional for any
$x\in\P^{17}$). Thus it only remains to show that $f(R(x,n)) =
R(x,n+1).$ While this brute force approach must, in principle,
lead to a straightforward proof of the theorem, the difficulty
lies in the considerable complexity and the high dimensionality of
the polynomial dynamical system
(\ref{eq:explicitPolynomialDynamicalSystem}). In fact, the first
component of the vector $f(R(x,2))$ is the square of a polynomial
of degree 4 with 110\,490 monomials. Other components of this
vector are at least as complex as the first one. While modern
supercomputers theoretically allow one to deal with polynomials of
this size, it is a task of formidable computational complexity to
carry out such a calculation. The author's attempts to perform it
on Nvidia Tesla M2090 supercomputer platform with a peak
performance of 16.872~Tflops (Linpack tested) were all
unsuccessful. Besides, it would not provide any explanation for
how~(\ref{eq:bloodDistrAfterNGenerations}) arose. For these
reasons we will follow a different way of proving
Theorem~\ref{thm:mainTheorem}.
\end{remark}

\begin{proof} Throughout the proof, we will be working with
projective coordinates of blood genotypes distributions. Thus all
the equalities below will relate vectors in projective spaces
meaning that two nonzero vectors are equal if and only if they are
proportional. Let $x=(x_1,x_2,\ldots,x_{18})$ $\in\P^{17}$ be the
vector encoding the initial blood genotypes distribution in a
population. It is straightforward to check that the action of the
polynomial map~(\ref{eq:explicitPolynomialDynamicalSystem}) on $x$
is given by
\begin{equation}
f(x) = (f_1(x),\ldots,f_{18}(x)) = Q(M(x)) =
Q(M_1(x),\ldots,M_6(x)). \label{eq:PolynomialDynamicalSystem}
\end{equation}
Here $M_i = M_i(x)$ is the $i$-th component of the image of the
initial distribution $x\in\P^{17}$ under the linear map $M:\P^{17}
\rightarrow \P^{5}$ defined by the matrix~(\ref{matrixM}).

We further observe that the matrix~$M$ has tensor product
structure:
\begin{equation}
\phantom{---------} M = \left(
\begin{array}{cccccc}
2 & 1 & 0 & 1 & 0 & 0   \\
0 & 1 & 2 & 0 & 0 & 1   \\
0 & 0 & 0 & 1 & 2 & 1
\end{array}
\right) \otimes \left(
\begin{array}{ccc}
2 & 1 & 0  \\
0 & 1 & 2
\end{array}
\right). \label{eq:tensorProductDecomposition}
\end{equation}
This equality is the algebraic counterpart of the blood genotypes
inheritance rule stating that the blood group variation and the Rh
factor variation are inherited independently. We define the map
$T:\P^5 \rightarrow \P^5$ to be the composition of the linear map
$M$ and the quadratic map $Q$ defined by (\ref{mapQ}) in the
reversed order: $T(y) = M(Q(y)).$ Using
(\ref{eq:tensorProductDecomposition}) we conclude that the $n$-th
iterate $T^n$ of the quadratic map $T$ acts on $y\in \P^5$ in
accordance with the formula {\small
\begin{equation}
\begin{array}{l}
T^n(y) =\phantom{---------------------------} \\
\phantom{T^n} \left(\right. 2^n y_1^2 + 2^n y_1 y_2 + 2^n y_1 y_3
+ 2^n y_1 y_4 + (2^n - 1)
y_2 y_4 + (2^n - 1) y_3 y_4 + y_1 y_5 + y_1 y_6, \\
\phantom{T^n} 2^n y_1 y_2 + 2^n y_2^2 + 2^n y_2 y_3 + y_2 y_4 +
(2^n - 1) y_1
y_5 + 2^n y_2 y_5 + (2^n - 1) y_3 y_5 + y_2 y_6, \\
\phantom{T^n} 2^n y_1 y_3 + 2^n y_2 y_3 + 2^n y_3^2 + y_3 y_4 +
 y_3 y_5 + (2^n - 1) y_1 y_6 + (2^n - 1) y_2 y_6 + 2^n y_3 y_6, \\
\phantom{T^n} 2^n y_1 y_4 + y_2 y_4 + y_3 y_4 + 2^n y_4^2 + (2^n -
1) y_1 y_5 +
 2^n y_4 y_5 + (2^n - 1) y_1 y_6 + 2^n y_4 y_6, \\
\phantom{T^n}  (2^n - 1) y_2 y_4 + y_1 y_5 +
 2^n y_2 y_5 + y_3 y_5 + 2^n y_4 y_5 + 2^n y_5^2 + (2^n - 1) y_2 y_6 +
 2^n y_5 y_6, \\
\phantom{T^n} (2^n - 1) y_3 y_4 + (2^n - 1) y_3 y_5 + y_1 y_6 +
y_2 y_6 +
 2^n y_3 y_6 + 2^n y_4 y_6 + 2^n y_5 y_6 + 2^n y_6^2 \left.\right).
\end{array}
\label{eq:MAndQinReversedOrder}
\end{equation}
}
It follows from~(\ref{eq:PolynomialDynamicalSystem}) that
$$
f^{n}(x) = (Q\circ M)^{n}(x) = Q\left( (M\circ Q)^{n-1} \left(
M(x) \right) \right) = Q\left( T^{n-1} (M(x)) \right).
$$
Recalling that $y = M(x)$ and
using~(\ref{eq:MAndQinReversedOrder}) we arrive
at~(\ref{eq:bloodDistrAfterNGenerations}). This finishes the
proof.
\end{proof}

Recall that we identify proportional distributions,
so~(\ref{eq:bloodDistrAfterNGenerations}) can be divided by the
sum of its components or any other normalizing common factor. The
following statement is an immediate consequence of
Theorem~\ref{thm:mainTheorem}.

\begin{corollary}
Blood genotypes frequencies after infinitely many generations are
obtained by passing to the limit
in~(\ref{eq:bloodDistrAfterNGenerations}). They are given by the
tensor product of the Hardy-Weinberg equilibrium frequencies of
the blood groups variations and Rh factor variations. These
frequencies span an attracting invariant manifold of the dynamical
system (\ref{eq:explicitPolynomialDynamicalSystem}).
\end{corollary}

\begin{remark}\rm
Describing evolutionary trajectories of initial distributions of
genotypes frequencies with respect to a given set of traits is a
classical avenue of research in population genetics. While
Theorem~\ref{thm:mainTheorem} is a formal consequence of the
Lyubich general evolution formula, see~\S~11 in
\cite{Lyubich1971}, few cases admit explicit description.
\end{remark}

\begin{example}\label{ex:exampleWithTable} \rm To illustrate
the action of the dynamical system
(\ref{eq:explicitPolynomialDynamicalSystem}) on a simple initial
distribution, we begin by treating a distribution spanned by two
genotypes. Consider an initial population where the first Rh
negative blood group is found with frequency~$0<p<1,$ the fourth
Rh homozygously positive blood group is found with frequency $1-p$
while no other blood group and Rh factor variations are present.
Table~1 summarizes the evolution of this distribution (empty cells
stand for zero frequencies).

\medskip

\begin{landscape}

\noindent
\begin{center}
\begin{minipage}{17cm}
\centerline{\bf \small Table~1: Blood group and Rh factor
phenotypes frequencies evolution} \centerline{\bf \small of the
initial population $p\cdot$OOhh + $(1-p)\cdot$ABHH}
\label{tab:bloodGroupsDistributions} {\tiny \vskip0.2cm \noindent
\begin{tabular}{|m{1.9cm}|m{0.7cm}|m{1.9cm}|m{2cm}|m{2cm}|m{2.9cm}|m{1.9cm}|}
\hline
{\bf Blood va\-ria\-tion} & Initial gen.& 1st generation & 2nd generation & 3rd generation & $n$-th generation \mbox{($n\geq 2$)} & After in\-fi\-ni\-te\-ly ma\-ny ge\-ne\-ra\-ti\-ons  \\
\hline O              & $p$ & $p^2$           & $p^2$    & $p^2$    & $p^2$      & $p^2$   \\
\hline A              &     & $(1+2p-3p^2)/4$ & $(1+2p-3p^2)/4$  & $(1+2p-3p^2)/4$     &  $(1+2p-3p^2)/4$   & $(1+2p-3p^2)/4$   \\
\hline B              &     & $(1+2p-3p^2)/4$ & $(1+2p-3p^2)/4$  & $(1+2p-3p^2)/4$     &  $(1+2p-3p^2)/4$   & $(1+2p-3p^2)/4$   \\
\hline AB             &$1-p$& $(1-p)^2/2$     & $(1-p)^2/2$  &  $(1-p)^2/2$    &  $(1-p)^2/2$   &  $(1-p)^2/2$  \\
\hline
\hline Rh positive    &$1-p$& $1-p^2$         & $1-p^2$  & $1-p^2$     &  $1-p^2$   & $1-p^2$   \\
\hline Rh negative    & $p$ & $p^2$           & $p^2$  & $p^2$     &  $p^2$   & $p^2$   \\
\hline
\hline O, Rh positive &     &                 & $(1-p)(3+p)\cdot p^2/4$  &  $(5-2 p-3 p^2)\cdot 3 p^2/16$    &  $4^{-n} (2^n-2) (1-p) p^2\cdot (2+2^n+(2^n-2) p)$   & $(1-p) p^2 (1+p)$    \\
\hline A, Rh positive &     & $(1+2p-3p^2)/4$ & $(1-p)(4+12 p-5 p^2-3 p^3)/16$ & $(1-p)(16+48 p-21 p^2-27 p^3)/64$ & $4^{-n-1} (1-p)\cdot(4^n-p(-3\cdot 4^n+(2^n-2)p(6+2^n+3(2^n-2) p)))$ & $ (1-p)^2 (1+p)\cdot(1+3 p)/4$   \\
\hline B, Rh positive &     & $(1+2p-3p^2)/4$ & $(1-p)(4+12 p-5 p^2-3 p^3)/16$ & $(1-p)(16+48 p-21 p^2-27 p^3)/64$ & $4^{-n-1} (1-p)\cdot(4^n-p(-3\cdot 4^n+(2^n-2)p(6+2^n+3(2^n-2) p)))$ & $(1-p)^2(1+p)\cdot(1+3p)/4$ \\
\hline AB, Rh positive&$1-p$& $(1-p)^2/2$     & $(1-p)^2 (4-p^2)/8$ & $(1-p)^2 (16-9 p^2)/32$ & $(4^n-(2^n-2)^2 p^2)\cdot 2^{-2n-1} (1-p)^2$ & $(1-p)^2(1-p^2)/2$ \\
\hline O, Rh negative & $p$ & $p^2$           & $p^2 (1+p)^2/4$ & $p^2 (1+3 p)^2/16$ & $4^{-n}p^2(2+(2^n-2)p)^2$ &  $p^4$  \\
\hline A, Rh negative &     &                 & $(1-p)(5+3p)\cdot p^2/16$ & $(1-p)(7+9 p)\cdot 3p^2/64$ & $4^{-n-1}(2^n-2)(1-p)\cdot p^2(6+2^n+3(2^n-2)p)$ & $(1-p)(1+3p)\cdot p^2/4$ \\
\hline B, Rh negative &     &                 & $(1-p)(5+3p)\cdot p^2/16$ & $(1-p)(7+9 p)\cdot 3p^2/64$ & $4^{-n-1}(2^n-2)(1-p)\cdot p^2(6+2^n+3(2^n-2)p)$ & $(1-p)(1+3p)\cdot p^2/4$ \\
\hline AB, Rh negative&     &                 & $(1-p)^2 p^2/8$ & $9(1-p)^2 p^2/32$ & $(2^n-2)^2(1-p)^2\cdot 2^{-2n-1}p^2$ & $(1-p)^2 p^2/2$ \\
\hline
\hline All blood types&  1  &       1         & 1  &  1  &  1   &  1  \\
\hline
\end{tabular}
}
\end{minipage}
\end{center}

\end{landscape}

\medskip

Table~1 shows that, although the blood groups and Rh factor
frequencies alone stabilize after one generation, the frequency of
the blood with any given combination of blood group and Rh factor
phenotypes (such as OH) does not remain constant after any finite
number of generations. There exists however the limit distribution
of frequencies that is achieved after infinitely many generations
and is described by the Hardy-Weinberg equilibrium. In fact, the
limit frequencies of the 8 phenotypes are given by the tensor
product of the frequencies of the four blood groups and the two
values of Rh factor.
\end{example}

\begin{example}\rm To illustrate the nontrivial dynamics of a
generic initial distribution of blood genotypes frequencies, we
consider the evolution of the distribution 2$\cdot$OOhh + AOhh +
2$\cdot$ABHH spanned by three genotypes. The phenotypes
frequencies in the subsequent generations are shown in
Fig.~\ref{fig:frequenciesDynamics}.

\begin{figure}[ht!]
\begin{center}
\begin{minipage}{12cm}
\centering
\includegraphics[width=12cm]{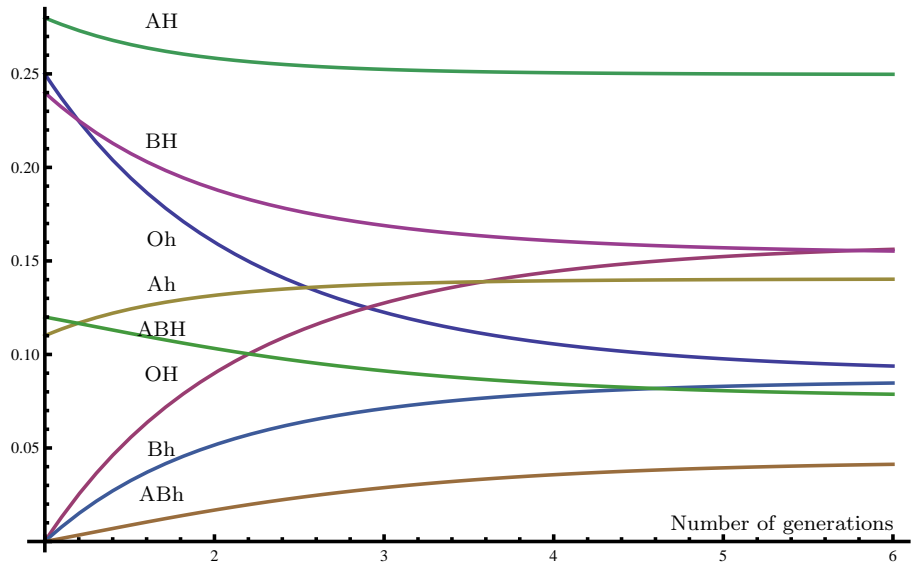}
\caption{\small Frequencies of the human blood phenotypes for the
initial distribution of genotypes 2$\cdot$OOhh + AOhh +
2$\cdot$ABHH} \label{fig:frequenciesDynamics} \vskip-9.4cm
 \hskip -8cm {\tiny AH} \vskip1cm \hskip-8cm {\tiny BH}
\vskip0.7cm \hskip-8cm {\tiny Oh} \vskip0.0cm \hskip-8.0cm {\tiny
Ah} \vskip-0.0cm \hskip-8cm {\tiny ABH} \vskip0.0cm \hskip-8cm
{\tiny OH} \vskip0.4cm \hskip-8cm {\tiny Bh} \vskip0.0cm
\hskip-8cm {\tiny ABh} \vskip-0.2cm \hskip8.5cm {\tiny Number of
generations} 
\end{minipage}
\end{center}
\end{figure}

\vskip3cm

We remark that it takes six human generations (around 200 years in
the real time scale) for the phenotypes frequencies in this
example to arrive at the 2\% relative error neighborhood of their
limit values.
\end{example}


\section{Mathematica package for the analysis of the blood groups frequencies}

While the evolutionary trajectory of any particular blood
genotypes distribution is completely described by
Theorem~\ref{thm:mainTheorem}, the algebraic structure of the
attracting invariant manifold of the dynamical
system~(\ref{eq:explicitPolynomialDynamicalSystem}) is far from
being clear. The analysis of its properties is a task of
substantial computational complexity and was done by means of a
package developed by the author and run under computer algebra
system {\it Mathematica}~9.0. One of the core algorithms
implemented in this package is as follows.

\medskip

\noindent{\bf Algorithm~1.}

\texttt{Step 0} \quad Define the list~\texttt{G} with the~18 blood
genotypes names OOhh, \ldots, ABHH as formal symbolic
algebraically independent variables. A population will from now on
be identified with a linear form in the elements of~\texttt{G}.
Its mass is defined to be the sum of the coefficients of this
linear form. Denote the space of all such forms by~\texttt{L(G)}.

\texttt{Step 1} \quad Define the blood genotypes inheritance
matrix~\texttt{A} to be the matrix of normalized (i.e., with
mass~1) linear forms in the elements of \texttt{G} that encode the
human blood genotypes inheritance rules.

\texttt{Step 2} \quad Define a bilinear map~\texttt{S} acting
on~\texttt{G}$\times$\texttt{G} and with values in \texttt{L(G)}
by means of the matrix~\texttt{A}. With this map, the normalized
next generation is computed as follows:

\texttt{NextGeneration[population\_]:=\newline \indent
Collect[Simplify[Expand[S[population,population]]/ \newline
\indent mass[population]],G]}.

\texttt{Step 3} \quad Choose a population by specifying the values
of some of the elements of the list~\texttt{G} and imposing
algebraic relations on the other.

\texttt{Step 4} \quad Find an algebraic parametrization of the
attracting submanifold for the population in question by
integrating the evolutionary equations.

\texttt{Step 5} \quad Eliminate the parameters and return the
complete set of algebraic equations defining the attracting
submanifold.

\medskip

While there exist several computer programs for the numerical
simulation of the evolution of recombination frequencies, the
above algorithm appears to be new. The structure of invariant
manifolds of a general multivariate map defined by a family of
quadratic forms is far from being clear and presumably does not
admit any algebraic description. The linearization of an
evolutionary trajectory in a neighborhood of the equilibrium
manifold has been given in \cite{Lyubich1971}.

Computer experiments with the {\it Mathematica} package reveal the
following intrinsic property of the attracting manifold: it is
given by a {\it binomial ideal} \cite{Eisenbud-Sturmfels}
generated by quadratic forms. A full list of these forms (many of
them being algebraically dependent) contains~96 elements and the
following shape: $ x_1 x_{10} = x_4 x_7, \,\, x_1 x_{11} = x_5
x_7, \,\, x_1 x_{12} = x_6 x_7, \,\, 4 x_1 x_{13} = x_{7}^2, 4
x_{1} x_{14} = x_{7} x_{8},  \,\, x_{10} x_{14} = 2 x_{12}
x_{13},$ $16 x_{1} x_{15} = x_{8}^2, \ldots, 4 x_7 x_9 = x_{8}^2.
$

We now apply Algorithm~1 to investigate invariant manifolds of the
polynomial dynamical
system~(\ref{eq:explicitPolynomialDynamicalSystem}). We consider
special cases of particularly simple distributions of blood
genotypes in the initial population.

\subsection{First blood group}

Consider the special case of a population consisting of people
with first blood group only (such as present day's south american
indians, see \cite{Hoffman2012}, p.~2189, Table~132-2). Assume
that the population in question comprises $a$ people with the
genotype OOhh, $b$ people with the genotype OOHh, and $c$ people
with the genotype OOHH. Using the notation introduced above, we
will denote such a population by $a\cdot {\rm OOhh} + b\cdot {\rm
OOHh} + c\cdot {\rm OOHH}.$ Then, in accordance with the blood
inheritance rules and their mathematical
formulation~(\ref{eq:explicitPolynomialDynamicalSystem}), the
blood genotypes distribution in the next generation will be
\begin{equation}
(2 a + b)^2 \cdot {\rm OOhh} +  2 (2 a + b) (b + 2 c) \cdot {\rm
OOHh} +  (b + 2 c)^2 \cdot {\rm OOHH}
\label{eq:limitDistrFirstGroup}
\end{equation}
and it will remain unchanged in any subsequent generation. In
other words, the variety parametrized for $(a,b,c)\in\P^2$ by
{\small
$$
\left( (2 a + b)^2, 0,0,0,0,0, 2 (2 a + b) (b + 2 c), \right.
\left. 0,0,0,0,0, (b + 2 c)^2, 0,0,0,0,0 \right) \in \P^{17}
$$
}
is an invariant manifold of the polynomial
map~(\ref{eq:explicitPolynomialDynamicalSystem}) which moreover
consists of fixed points of this map. The three nonzero
equilibrium frequencies $x_1,x_7,x_{13}$ of the three genotypes in
this example lie on the discriminant hypersurface $4 x_1 x_{13} =
x_{7}^2.$

From now on we will be using the linear form notation for blood
genotypes distributions since they allow one to avoid vectors with
plenty of zeros.

\subsection{Rh negative blood}

Since Rh negative blood is a recessive trait, a population
consisting of Rh negative people only can be represented in the
following form:
$$
p \cdot {\rm OOhh} + q \cdot {\rm AOhh} + r \cdot {\rm AAhh} + u
\cdot {\rm BOhh} + v \cdot {\rm BBhh} + w \cdot {\rm ABhh}.
$$
Such a distribution of blood genotypes will also stabilize already
in the next generation. This new stable distribution is given by
\begin{equation}
\begin{array}{c}
(2 p+q+u)^2 \cdot {\rm OOhh} + 2 (2 p+q+u) (q+2 r+w) \cdot {\rm
AOhh} + \\
(q+2 r+w)^2 \cdot {\rm AAhh} + 2 (2 p+q+u) (u+2 v+w) \cdot {\rm
BOhh} + \\
(u+2 v+w)^2 \cdot {\rm BBhh} +  2 (q+2 r+w) (u+2v+w)
\cdot {\rm ABhh}.\phantom{-}
\end{array}
\label{eq:limitDistrRhNegative}
\end{equation}
For any $(p,q,r,u,v,w)\in\P^5$ the point
(\ref{eq:limitDistrRhNegative}) is another fixed point of the
polynomial map~(\ref{eq:explicitPolynomialDynamicalSystem}). It is
easily seen that the equilibrium frequencies satisfy the binomial
relations $4 x_1 x_3 = x_2^2 , \,\, 4 x_1 x_5 = x_4^2 , \,\, 2 x_2
x_5 = x_4 x_6 , \,\, 4 x_3 x_5 = x_6^2 , \,\, 2 x_1 x_6 = x_2 x_4
, \,\, x_2 x_6 = 2 x_3 x_4.$

\subsection{Evolution of the population $a\cdot$OOhh + $b\cdot$AAhh +
$c\cdot$AAHH}

Consider the population $P = a\cdot$OOhh + $b\cdot$AAhh +
$c\cdot$AAHH. Here $a,b$ and $c$ are arbitrary positive numbers
representing proportions of people with corresponding blood
genotypes. Any real population is of course very far from having
such a distribution of blood genotypes. We consider this example
since it is essentially different from the previous ones and still
allows one to explicitly compute the limit distribution of the
blood genotypes.

Already the third generation of the population~$P$ defined above
will contain all nine genotypes that belong to the first or the
second blood group. Using the Wolfram Mathematica~9.0 computer
algebra system and a package for blood genotypes analysis we
conclude that after sufficiently many generations the blood
genotypes distribution in the population under study will be
arbitrarily close to the limit distribution
\begin{equation}
\begin{array}{r}
a^2 (a+b)^2 \cdot {\rm OOhh}  +  2  a (a+b)^2 (b+c)  \cdot {\rm
AOhh} +  (a+b)^2(b+c)^2 \cdot  {\rm AAhh}  + \\
2  a^2 (a+b) c \cdot{\rm OOHh}  + 4 a (a+b) c (b+c)  \cdot {\rm
AOHh}  +
2 (a+b) c (b+c)^2  \cdot {\rm AAHh} + \\
a^2 c^2  \cdot  {\rm OOHH} +  2  a c^2 (b+c)  \cdot {\rm AOHH} +
c^2 (b+c)^2  \cdot {\rm AAHH}.\phantom{..}
\end{array}
\end{equation} For any initial frequencies $a,b,c$ this
blood genotypes distribution is invariant under the map~$f.$ The
nonzero equilibrium frequencies span the manifold defined by the
binomial equations $4 x_1 x_{13} = x_{7}^2, \,\, 4 x_{1} x_{14} =
x_7 x_8, \ldots,$ $4 x_7 x_9 = x_8^2.$

\subsection{Blood with the same distribution of blood groups for all Rh
genotypes}

Statistics shows that in most populations blood group and Rh
factor distributions do not correlate with each other
\cite{Hoffman2012,Kang1997,Ohashi2006,Sato2010}\,. Since blood
genotype (including all the information on homo- or heterozygosity
of a person for blood group and Rh factor) is much more difficult
to detect clinically than the dominating blood group and Rh
factor, the corresponding statistics for their variations is not
available. Yet, genetics of these traits suggests to consider them
as statistically independent. In the present example, we
investigate the blood genotypes dynamics of a population
satisfying this additional assumption.

Such a population is completely determined by the two vectors $R =
(a,b,c)\in\P^2$ and $G = (p,q,r,u,v,w)\in\P^5$ giving the numbers
of people with Rh-factor variations (hh,Hh,HH) and the
distribution of blood group variations (OO, AO, AA, BO, BB, AB)
within every such set. With this notation, the 18-dimensional
vector defining a population that satisfies the above assumption
is given by the tensor product of the vectors $R$ and $G$ defined
as follows: $ R\otimes G = (a p, a q, a r, a u, a v, a w, b p, b
q, b r, b u, b v, b w, c p, c q,$ $c r, c u, c v, c w). $
Computation shows that the blood genotypes distribution of such a
population will also stabilize in the next generation and the new
stable distribution is the tensor product of the
distributions~(\ref{eq:limitDistrFirstGroup})
and~(\ref{eq:limitDistrRhNegative}):
$$
{\small
\begin{array}{c}
 \left( \right. A (2 p+q+u)^2, \,\,
 2 A (2 p+q+u) (q+2 r+w), \,\,
 A (q+2 r+w)^2, \,\,
 2 A (2 p+q+u) (u+2 v+w), \\
 A (u+2 v+w)^2, \,\,
 2 A (q+2 r+w) (u+2 v+w), \,\,
 B (2 p+q+u)^2, \,\,
 2B (2 p+q+u) (q+2 r+w), \\
 B (q+2 r+w)^2, \,\,
 2B (2 p+q+u) (u+2 v+w), \,\,
 B (u+2 v+w)^2, \,\,
 2B (q+2 r+w) (u+2 v+w), \\
 C (2 p+q+u)^2, \,\,
 2 C (2 p+q+u) (q+2 r+w), \,\,
 C (q+2 r+w)^2, \,\,
 2 C (2 p+q+u) (u+2 v+w), \\
 C (u+2 v+w)^2, \,\,
 2 C (q+2 r+w) (u+2 v+w) \left. \right),
\end{array}
}
$$
where $A=(2 a+b)^2,$  $B = 2 (2 a+b) (b+2 c),$ $C = (b+2 c)^2.$


\section{Back to real data}
\label{sec:discussion}

The image of the space of all blood genotypes distributions under
the map~(\ref{eq:explicitPolynomialDynamicalSystem}) encoding the
blood inheritance rules is six-dimensional. Thus the evolution of
an 18-dimensional initial distribution is completely determined by
its six parameters which can be chosen to be the frequencies of
the fully homozygous blood genotypes. The initial distributions
that evolve differently are those that do not differ by a vector
in the kernel of the matrix~(\ref{matrixM}).

\begin{figure}[ht!]
\begin{center}
\begin{minipage}{12cm}
\centering
\includegraphics[width=12cm]{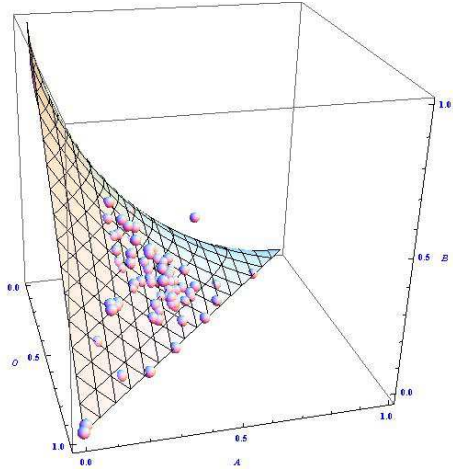}
\caption{\small The Bernstein hypersurface $1+\sqrt{{\rm
O}}=\sqrt{{\rm A+O}}+\sqrt{{\rm B+O}}$ and the observed
distributions of blood groups frequencies in the world}
\label{fig:BernsteinHypersurface}
\end{minipage}
\end{center}
\end{figure}

It is classically known that the equilibrium phenotypes
frequencies O,A and B of the 1st, 2nd and 3rd blood groups
respectively satisfy the Bernstein algebraic relation
\cite{BernsteinFelix1929,Novitski1976}\, $1+\sqrt{{\rm
O}}=\sqrt{{\rm A+O}}+\sqrt{{\rm B+O}}.$ This relation (together
with O+A+B+AB=1) allows one to express e.g. the frequency of the
4th blood group as a function $F_{{\rm AB}}({\rm O,A})$ of the
frequencies of the first two. Using statistical data available at
www.bloodbook.com/world-abo.html where blood group distributions
for~88 ethnicities of the world are collected, we find that the
relative error of the estimate $F_{{\rm AB}}({\rm O,A})$ exceeds
30\% in 11\% of all cases. Also, this relative error is greater
than 20\% for 19\% of all the observed blood groups phenotypes
distributions. This suggests that many of the observed
distributions do not lie on the Bernstein hypersurface (see
Fig.~\ref{fig:BernsteinHypersurface}) and therefore are not at the
equilibrium and will necessarily evolve further.
The formula~(\ref{eq:bloodDistrAfterNGenerations}) gives a
complete description of their evolutionary trajectories in the
absence of evolutionary influences like meiosis, migration etc.

While extensive statistical data on blood phenotypes distributions
in the various populations of the world is available at
www.bloodbook.com/world-abo.html and similar sources, little is
known about the human blood genotypes distributions. The reason
for this is that a human's blood genotype is much more difficult
to detect clinically than her/his blood phenotype. However, to
compute the expected blood phenotypes or genotypes distribution in
the next generation, the present genotypes distribution must be
known. This lack of statistical data does not allow us to directly
apply Theorem~\ref{thm:mainTheorem} to a real population. Yet, the
formula~(\ref{eq:bloodDistrAfterNGenerations}) shows which initial
distributions evolve along the same trajectories as well as their
rate of convergence towards the equilibrium.



\begin{thebibliography}{}

\bibitem{Anstee2010}
Anstee, D.J., 2010. The relationship between blood groups and
disease. { Blood} {\bf 115}, 4635-4643.

\bibitem{Kang1997}
Kang S.H. et al., 1997. Distribution of abo genotypes and allele
frequencies in a korean population. { Japanese Journal of Human
Genetics} {\bf 42}, 331-335.

\bibitem{Ohashi2006}
Ohashi J. et al., 2006. Polymorphisms in the ABO blood group gene
in three populations in the New Georgia group of the Solomon
Islands. { Journal of Human Genetics} {\bf 51}, 407-411.

\bibitem{Sato2010}
Sato T. et al., 2010. Polymorphisms and allele frequencies of the
ABO blood group gene among the Jomon, Epi-Jomon and Okhotsk people
in Hokkaido, northern Japan, revealed by ancient DNA analysis. {
Journal of Human Genetics} {\bf 55}, 691-696.

\bibitem{Novitski1976}
Novitski E., 1976. ABO blood groups and the Hardy-Weinberg
equilibrium. { Science} {\bf 6}, 478.

\bibitem{Bernstein1923}
Bernstein, S.N., 1923. Principe de stationarite et generalisation
de la loi de Mendel. { C.R. Acad. Sci. Paris} {\bf 177}, 528-531.

\bibitem{BernsteinFelix1929}
Bernstein, F., 1930. \"Uber die Erblichkeit der Blutgruppen. {
Zeitschrift f\"ur Induktive Abstammungs- und Vererbungslehre} {\bf
54}:1, 400-426.

\bibitem{Lyubich1971}
Lyubich Y.I., 1971. Basic concepts and theorems on the
evolutionary genetics of free populations. { Russian Mathematical
Surveys} {\bf 26}:5, 51-123.

\bibitem{Hoffman2012}
Hoffman R. et al., 2012. { Hematology: Basic Principles and
Practice} (6th ed.). Elsevier. ISBN 978-1-4377-2928-3.

\bibitem{Okada}
Okada Y., Kamatani Y., 2012. Common genetic factors for
hematological traits in Humans. { Journal of Human Genetics} {\bf
57}, 161-169.

\bibitem{Reilly2007}
Reilly M., Szulkin R., 2007. Statistical analysis of
donation-transfusion data with complex correlation. { Stat. Med.}
{\bf 26}:30, 5572-5585.

\bibitem{BedfordJonsson2000}
Bedford, E., Jonsson, M., 2000. Dynamics of regular polynomial
endomorphisms of $\C^k$. { Amer. J. Math.} {\bf 122}:1, 153-212.

\bibitem{Cantat2010}
Cantat S., Chambert-Loir A., Guedj V., 2010. { Quelques aspects
des syst\`{e}mes dynamiques polynomiaux}. (French) [{ Some Aspects
of Polynomial Dynamical Systems}.] Panoramas et Synth\`{e}ses
[Panoramas and Syntheses], 30. Soci\'{e}t\'{e} Math\'{e}matique de
France, Paris. x+341 pp. ISBN: 978-2-85629-338-6.

\bibitem{Eisenbud-Sturmfels}
Eisenbud D., Sturmfels B., 1996. Binomial ideals. Duke Math.~J.
{\bf 84}:1, 1-45.

\end{thebibliography}

\end{document}